\documentclass[twocolumn,aps,pra]{revtex4-2}
\usepackage{bm,bbm}
\usepackage{amsfonts,amsmath,amssymb,latexsym}
\usepackage{graphicx}
\usepackage{mathtools}
\usepackage{xcolor}
\usepackage{hyperref}
\begin{document} 
\title{Backflow in vector Gaussian beams}
\author{Tomasz Rado\.zycki}
\email{t.radozycki@uksw.edu.pl}
\affiliation{Faculty of Mathematics and Natural Sciences, College of Sciences, Institute of Physical Sciences, Cardinal Stefan Wyszynski University in Warsaw, W\'oycickiego 1/3, 01-938 Warsaw, Poland} 

\begin{abstract}
The phenomenon of Poynting backflow in a single vector Gaussian beam is examined. The paraxial Maxwell equations and their exact solutions containing terms proportional to the small parameter $\varepsilon=\frac{\lambda}{2\pi w_0}$, or to its square, where $w_0$ is the beam waist, are made use of. Explicit analytical calculations show that just these additional expressions are responsible for the occurrence of the reversed Poynting-vector longitudinal component for selected polarizations. Concrete results for the time-averaged vector for several Gaussian beams with $n=1$ ($n$ being the topological index of the beam) are presented. Depending on the choice of polarization, the backflow area is located around the beam's axis, has an annular character or is absent. In the case of $n=2$, backflow areas were found as well. The magnitude of the backflow proved to be proportional to $\varepsilon^4$.
\end{abstract}

\maketitle

\section{Introduction}\label{intro}

Backflow is a wave phenomenon in which the probability current or Poynting vector exhibits a flow in the opposite direction than would be inferred from the components of momentum present	 in the spectral decomposition of the wave. The interest in this phenomenon dates back to the 1960s and the works related to the arrival time of a quantum particle \cite{al1,al2,al3}.
In \cite{brack} the authors analyzed the backflow in one-dimensional quantum mechanics (QM) by examining the probability current $j(t,x)$. The normalized solution of the Schr\"odinger equation comprising in the spectrum only positive momenta (velocities), generated a probability current, which for a certain period of time showed a flow in the direction opposite to that exhibited by the momentum components. In particular, it turned out that $j(0,0)<0$ or even $j(t,0)<0$ over a certain time interval. By virtue of the scaling argument, this interval can be made arbitrarily long, at the price of a suitable reduction in the instantaneous value of the backflow. In the quoted work the maximal increase of the probability on the ``wrong'' side of zero (i.e. the so called backflow constant $\lambda$) was established approximately at $0.04$ . 

The same effect was demonstrated for the two-component Dirac wave-function in one dimension \cite{mell,asfa} and the non-relativistic value of $\lambda$ was more precisely set at $0.039$ (this value was confirmed -- or even made still more accurate in \cite{penz}). The effect persists even for a particle -- i.e. the wave-packet -- subject to a constant force positively directed, although it obviously weakens with the increasing value of the force \cite{mel2}.

In \cite{berry}, a thorough analysis of the effect was carried out, including both the time evolution of the backflow regions in one-dimensional QM and their distribution in a two-dimensional paraxial optical beam. In the latter case, the backflow fraction turned out to be proportional to $\theta_{\mathrm{max}}^4$, where $\theta_{\mathrm{max}}$ stands for the maximum angle of inclination of the plane wave component in the beam. This quantity is roughly related to the parameter $\varepsilon$ defined by (\ref{deeps}), so this result coincides with those obtained in Sec. \ref{parro} in this work.

Subsequent works on QM have investigated the appearance of this effect in a superposition of waves \cite{saari} and also studied the classical limit $\hbar\to 0$ \cite{year1,year2}. A large family of backflow states in one-dimensional QM was given in \cite{hali}. It was also indicated that in more-than-one dimensions this phenomenon extends to angular momentum and rotational motion, such as that of an electron in a constant magnetic field \cite{strange,diba,ghg}.

In \cite{ibba}, it was pointed out that backflow is a widespread wave phenomenon (after all, ``quantum'' mechanics means the same as ``wave'' mechanics). Examples were given there of fully relativistic systems in three dimensions, such as Dirac and Maxwell waves, as well as exponential \cite{ibbc} and Bessel light beams, and even gravitational waves. 

The effect spoken of is not only a theoretical curiosity: it is attributed to interference phenomena, which should manifest themselves most directly in classical optics and actually have been experimentally observed \cite{elie,daniel}. Moreover, as written in \cite{daniel} they are ``surprisingly common'' in optical fields. It can have essential consequences for manipulating particles (as in optical tweezers \cite{ash,chu,miller,pad}) or controlling the dynamics of light beams.
In this paper we would like to demonstrate its appearance in the fundamental laser beam, namely the Gaussian beam \cite{sie, patta}, depending on the polarization. In \cite{kotab,kotac} on-axis backflow was found for the linearly polarized beam with vorticity $\pm 2$. In this paper, our intention is to show that if exact solutions of the paraxial Maxwell equations (including additional expressions proportional to $\varepsilon$ or $\varepsilon^2$ of \cite{trvec}) are used, backflow appears for a single Gaussian beam in various configurations (when two Gaussian beams are interfered with each other, the presence of retro-propagation has been observed experimentally in \cite{daniel}). 

It should be stressed, however, that there is a slight deviation from this concept if traditionally understood. Usually, as mentioned at the beginning, this notion is applied to the circumstance in which a wave packet constructed exclusively of components with positive momenta propagates partially in the opposite direction (for a quantum particle in one dimension, this requirement has been somewhat relaxed in \cite{millera,barb}). This was for instance the case of exponential and Bessel beams dealt with in \cite{ibba}. The Gaussian beam does not meet this condition: it is composed of negatively-valued wave vector components in the spectral distribution as well. This is quite clear: this beam is the solution of the Maxwell paraxial equations, and its characteristic feature is the non-trivial dependence on $z$ through the so-called complex beam parameter $1+i\frac{z}{z_{\mathrm{R}}}$. However, as it will be shown in Sec. \ref{vgb}, the contribution of those ingredients is exponentially small (proportional to $e^{-\frac{1}{2\varepsilon^2}}$) in contrast to the power-law backflow effect (as already mentioned $\sim\varepsilon^4$). The parameter $\varepsilon$ is usually small and remains {\em de facto} at our disposal, so in a typical situation the observed backflow amounts to many orders of magnitude more than would be inferred from the presence of negative components in the spectral distribution of the beam. 

In our work, we focus on the properties of the Poynting vector, leaving aside the discussion of the distinction between the Poynting current and the canonical current \cite{gogo}. The content is organized as follows. In Sec. \ref{moco} the paraxial Maxwell equations are provided together with their full solutions depending on two scalar ``potentials''. Sec. \ref{vgb} deals with the momentum spectral analysis of Gaussian beams. Finally, in Sec. \ref{parro}, a couple of examples exhibiting reversed Poynting vector are presented in analytical as well as graphical form.

\section{Vector paraxial fields}\label{moco}

We start with recalling the set of paraxial Maxwell equations derived in \cite{trvec}.
Since we are going to deal with monochromatic light beams propagating along the $z$ axis it is handy to temporarily extract the factor $e^{i(kz-\omega t)}$, and define the ``tilde'' quantities (e.g. complex electric and magnetic fields and so on) according to
\begin{equation}\label{monoch}
F(\bm{r},t)=e^{i(kz-\omega t)}\widetilde{F}(\bm{r}).
\end{equation}
Correspondingly the nabla operator acting on these ``tilde'' quantities will have the form
\begin{equation}\label{grab}
\bm{\nabla}=\left[\partial_x,\partial_y,ik+\partial_z\right],
\end{equation}
and its paraxial counterpart
\begin{equation}\label{pargrab}
\widetilde{\bm{\nabla}}=\left[\partial_x,\partial_y,ik+\frac{1}{2}\,\partial_z\right],
\end{equation}
Then the set of sourceless paraxial Maxwell equations for electromagnetic fields may be written as
\begin{subequations}\label{maxwellEB}
\begin{align}
\widetilde{\bm{\nabla}}\times\widetilde{\bm{E}}&=c\left(ik-\frac{1}{2}\,\partial_z\right)\widetilde{\bm{B}},\label{maxwellE}\\
\widetilde{\bm{\nabla}}\times\widetilde{\bm{B}}&=-\frac{1}{c}\left(ik-\frac{1}{2}\,\partial_z\right)\widetilde{\bm{E}},\label{maxwellB}\\
\widetilde{\bm{\nabla}}\widetilde{\bm{E}}&=0,\label{maxwellGE}\\
\widetilde{\bm{\nabla}}\widetilde{\bm{B}}&=0.\label{maxwellGB}
\end{align}
\end{subequations}
Bold symbols, depending on the context, refer either to true three-dimensional vectors, or to reduced two-dimensional ones (e.g. $\bm{\xi}$). Hopefully, this will not cause any confusion. 

As it was found in \cite{trvec} the general solutions to these equations can be expressed in terms of two scalar ``potentials'' $\widetilde{V}_\pm$ in the following way:
\begin{subequations}\label{parpeb}
\begin{align}
&\widetilde{E}_x=\left(1-\frac{i}{2k}\,\partial_z\right)\partial_x\widetilde{V}_+-i\left(1+\frac{i}{2k}\,\partial_z\right)\partial_y\widetilde{V}_-,\label{parpebx}\\
&\widetilde{E}_y=i\left(1+\frac{i}{2k}\,\partial_z\right)\partial_x\widetilde{V}_-+\left(1-\frac{i}{2k}\,\partial_z\right)\partial_y\widetilde{V}_+,\label{parpeby}\\
&\widetilde{E}_z=2\partial_z\widetilde{V}_+.\label{parpebz}
\end{align}
\end{subequations}
and 
\begin{subequations}\label{parpb}
\begin{align}
&\widetilde{B}_x=-\frac{1}{c}\left[i\left(1-\frac{i}{2k}\,\partial_z\right)\partial_x\widetilde{V}_-+\left(1+\frac{i}{2k}\,\partial_z\right)\partial_y\widetilde{V}_+\right],\label{parpbx}\\
&\widetilde{B}_y=\frac{1}{c}\left[\left(1+\frac{i}{2k}\,\partial_z\right)\partial_x\widetilde{V}_+-i\left(1-\frac{i}{2k}\,\partial_z\right)\partial_y\widetilde{V}_-\right],\label{parpby}\\
&\widetilde{B}_z=-\frac{2i}{c}\,\partial_z\widetilde{V}_-.\label{parpbz}
\end{align}
\end{subequations}
Both potentials can be arbitrarily chosen, except for the fact that they have to represent solutions of the scalar paraxial equation:
\begin{equation}\label{}
\left(\mathcal{4}_\perp+2ik\partial_z\right)\widetilde{V}_\pm=0,
\end{equation}
with $\mathcal{4}_\perp$ standing for the transverse, two-dimensional Laplace operator.

As it is well known any Gaussian beam has two characteristic length scales: $w_0$ -- the radius of the beam waist in the transverse plane, and $z_R=\frac{1}{2}\,kw_0^2$ -- the Rayleigh range along the propagation axis. It is highly comfortable to set them as measures of length and introduce corresponding dimensionless coordinates:
\begin{equation}\label{bezwb}
\xi_x=\frac{x}{w_0},\qquad \xi_y=\frac{y}{w_0},\qquad \zeta=\frac{z}{z_R}.
\end{equation}
The two-dimensional vector $[\xi_x,\xi_y]$ will be further denoted with $\bm{\xi}$. 

Since usually the following inequalities hold
\begin{equation}\label{defyb}
\lambda\ll w_0\ll z_R=\frac{kw_0^2}{2},
\end{equation}
the following dimensionless parameter appears to be helpful:
\begin{equation}\label{deeps}
\varepsilon=\frac{1}{kw_0}=\frac{\lambda}{2\pi w_0}=\frac{w_0}{2 z_R},
\end{equation}
This parameter can be really very tiny, as for instance $10^{-4}$ or so, but for tightly focused beams \cite{neice,luo,cheben,ren,kaz,gaon,kolbow,sharma,sted} it can reach values of the order of $0.2-0.3$.

In order to rewrite the formulas for the electromagnetic fields in the simplest possible and dimensionless form they will be redefined as follows:
\begin{equation}\label{bezeb}
\widetilde{\bm{\mathcal{E}}}=w_0 \widetilde{\bm{E}},\qquad \widetilde{\bm{\mathcal{B}}}=cw_0 \widetilde{\bm{B}}.
\end{equation}

With all these substitutions one gets the following expressions for the electric field
\begin{subequations}\label{parpeep}
\begin{align}
\;&\widetilde{\mathcal{E}}_{\xi_x}=\partial_{\xi_x}\left(1-i\varepsilon^2\partial_\zeta \right)\widetilde{V}_+-i\partial_{\xi_y}\left(1+i\varepsilon^2\partial_\zeta\right)\widetilde{V}_-,\label{parpexe}\\
&\widetilde{\mathcal{E}}_{\xi_y}=i\partial_{\xi_x}\left(1+i\varepsilon^2\partial_\zeta\right)\widetilde{V}_-+\partial_{\xi_y}\left(1-i\varepsilon^2\partial_\zeta\right)\widetilde{V}_+,\label{parpeye}\\
&\widetilde{\mathcal{E}}_{\zeta}=4\varepsilon\partial_\zeta\widetilde{V}_+,\label{parpeze}
\end{align}
\end{subequations}
and likewise for the magnetic field
\begin{subequations}\label{parpbeb}
\begin{align}
&\widetilde{\mathcal{B}}_{\xi_x}=-i\partial_{\xi_x}\left(1-i\varepsilon^2\partial_\zeta\right)\widetilde{V}_--\partial_{\xi_y}\left(1+i\varepsilon^2\partial_\zeta\right)\widetilde{V}_+,\label{parpbxe}\\
&\widetilde{\mathcal{B}}_{\xi_y}=\partial_{\xi_x}\left(1+i\varepsilon^2\partial_\zeta\right)\widetilde{V}_+-i\partial_{\xi_y}\left(1-i\varepsilon^2\partial_\zeta\right)\widetilde{V}_-,\label{parpbye}\\
&\widetilde{\mathcal{B}}_{\zeta}=-4i\varepsilon\partial_\zeta\widetilde{V}_-.\label{parpbze}
\end{align}
\end{subequations}
In the following sections, we will focus on Gaussian beams, by choosing particular forms of scalar functions $\widetilde{V}_\pm$. It will turn out that, in addition to that, the key role is played by the $\varepsilon^2$ terms in (\ref{parpeep}) and (\ref{parpbeb}).

\section{Spectrum of momenta}\label{vgb}

The fundamental Gaussian solution of the paraxial equation which in coordinates (\ref{bezwb}) reads
\begin{equation}\label{paraxz}
\left(\partial^2_{\xi_x}+\partial^2_{\xi_y}+4i\partial_\zeta\right)\widetilde{V}=0,
\end{equation}
has the form \cite{trmulti}
\begin{equation}\label{podfu}
\widetilde{V}_n(\bm{\xi},\zeta)=\frac{(\xi_x\pm i \xi_y)^n}{(1+i\zeta)^{n+1}}\,e^{-\frac{\bm{\xi}^2}{1+i\zeta}},
\end{equation}
or equivalently in cylindrical coordinates
\begin{equation}\label{podfuc}
\widetilde{V}_n(\xi,\phi,\zeta)=\frac{\xi^ne^{\pm in\phi}}{(1+i\zeta)^{n+1}}\,e^{-\frac{\xi^2}{1+i\zeta}}.
\end{equation}
These functions describe a standard scalar Gaussian beam carrying a vortex of topological charge (vorticity) $n$. Out of (\ref{podfu}) the electric and magnetic fields in various configurations can be generated via equations (\ref{parpeep}) and (\ref{parpbeb}).

In the backflow phenomenon it is essential to determine the spectrum of momenta. All this information is contained within expressions (\ref{podfu}) or  (\ref{podfuc}). Let us, therefore, calculate the Fourier transform of (\ref{podfuc}) with respect to $\zeta$:
\begin{equation}\label{tfov}
\widetilde{V}_n(\xi,\phi,q)=\int\limits_{-\infty}^{\infty}dq\, e^{-iq\zeta}\frac{\xi^ne^{\pm i\phi}}{(1+i\zeta)^{n+1}}\,e^{-\frac{\xi^2}{1+i\zeta}}.
\end{equation}
The only singularity (in fact it is the essential singularity) in the plane of complex $\zeta$ is located in the upper half-plane at the point $\zeta=i$. However, for positive values of $q$ the integration contour can be enclosed by the (infinite) semicircle lying in the lower half-plane, which leaves the singularity outside and leads to the vanishing of (\ref{tfov}). Therefore, one gets
\begin{eqnarray}
\widetilde{V}_n(\xi,\phi,q)&=&2\pi i\theta(-q)\xi^ne^{\pm i n\phi}\underset{\zeta=i}{\mathrm{Res}}\left[\frac{e^{-\frac{\xi^2}{1+i\zeta}-iq\zeta}}{(1+i\zeta)^{n+1}}\,\right]\nonumber\\
&=&2\pi i\theta(-q)\xi^ne^{\pm i n\phi}e^q\sum_{m=0}^\infty\frac{(-\xi^2)^m}{m!(1+i\zeta)^{m+n+1}}\nonumber\\
&&\times\sum_{l=0}^\infty\frac{[-q(1+i\zeta)]^l}{l!}\,\delta_{m+n,l}\label{vnfou}\\
&=&2\pi \theta(-q)(-q\xi)^ne^{\pm i n\phi}e^q\sum_{m=0}^\infty\frac{(q\xi^2)^m}{m!(m+n)!}\nonumber\\
&=&2\pi \theta(-q)(-q)^{n/2}e^{\pm i n\phi}e^qJ_n\left(2\xi\sqrt{-q}\right), \nonumber
\end{eqnarray}
where $J_n$ stands for the Bessel function, and consequently
\begin{eqnarray}
\widetilde{V}_n(\xi,\phi,\zeta)&=&e^{\pm in\phi}\int\limits_{-\infty}^0dq\,e^q(-q)^{n/2}J_n\left(2\xi\sqrt{-q}\right)e^{iq\zeta}\nonumber\\
&=&e^{\pm in\phi}\int\limits_0^{\infty}dq\,e^{-q}q^{n/2}J_n\left(2\xi\sqrt{q}\right)e^{-iq\zeta}.\label{pozaca}
\end{eqnarray}
This expression must be supplemented by the factor (\ref{monoch}), i.e. $e^{i\frac{\zeta-\tau}{2\varepsilon^2}}$, where $\tau=\frac{ct}{z_{\mathrm{R}}}$, so that
\begin{eqnarray}
V_n(\xi,\phi,\zeta,\tau)&=&e^{\pm in\phi}e^{-i\frac{\tau}{2\varepsilon^2}}\int\limits_0^{\infty}dq\,e^{-q}q^{n/2}\nonumber\\
&&\times J_n\left(2\xi\sqrt{q}\right)e^{i\left(\frac{1}{2\varepsilon^2}-q\right)\zeta}.\label{vbet}
\end{eqnarray}
As can be seen, this expression contains both positive (for $q<\frac{1}{2\varepsilon^2}$) and negative  (for $q>\frac{1}{2\varepsilon^2}$) momenta. However, it can be shown that the contribution from the latter ones is exponentially small, which is owed to the factor $e^{-q}$ in the integrand. The coefficient that determines the order of magnitude of this contribution, as will be shown below, is $e^{-\frac{1}{2\varepsilon^2}}$, which attains the value of $10^{-87}$ when $\varepsilon=0.05$, and reaches $3\times 10^{-4}$ for $\varepsilon=0.25$. For real beams, the negative momenta can be treated from the practical point of view as nonexistent.
In the appendix the role played by negative-valued momentum components is estimated in more precise manner. 

It stems from (\ref{refoi}) and (\ref{prmi}), that the contribution of the negative momenta to the whole packet is
\begin{equation}\label{contrn}
\left|\frac{e^{\pm in\phi}e^{-i\frac{\tau}{2\varepsilon^2}} {\cal I}_n}{V_n}\right|\sim C_n(\xi)\frac{(1+\zeta^2)^{n/2}}{\varepsilon^n}\,e^{-\frac{1}{2\varepsilon^2}},
\end{equation}
where
\begin{equation}\label{codx}
 C_n(\xi)=\frac{1}{2^{n/2+1/4}\pi^{1/2}\xi^n},
\end{equation}
or eventually for small $\xi$:
\begin{equation}\label{codnx}
 C_n(\xi)=\frac{1}{n!\varepsilon^n}.
\end{equation}
This means that the negative-momenta content in the Gaussian beam is actually exponentially small. In what follows we focus only on the beams with $n=1$ and $n=2$, so the presence of the additional power factor in denominator does not change this conclusion. Moreover, this holds even for larger values of $n$ if $\varepsilon$ is small enough (already for $\varepsilon\approx 0.1$ the value of the exponential factor drops to $10^{-22}$).

\section{Backflow in the Poynting vector}\label{parro}

In the present section we will demonstrate that for certain polarization configurations of the vector Gaussian beam the local backflow represented by the negative $\zeta$ component of the Poynting vector is observed, and its amount is much larger (in the elaborated examples the ratio is about $10^6$) than the content of the negative momenta in the spectrum of the beam.

\subsection{The $n=1$ case}\label{lipo}

Consider the Gaussian beam of order $n=1$ defined by the scalar function (\ref{podfu}), i.e.,
\begin{equation}\label{podfuop}
\widetilde{V}_1(\bm{\xi},\zeta)=\frac{\xi_x\pm i \xi_y}{(1+i\zeta)^2}\,e^{-\frac{\bm{\xi}^2}{1+i\zeta}}.
\end{equation}
This function satisfies the paraxial equation both for the `$+$' and `$-$' sign, and the same applies to their sum or difference with arbitrary coefficients since the paraxial equation is linear. Let us then define
\begin{subequations}\label{liparac}
\begin{align}
&\widetilde{V}_+(\xi_x, \xi_y,\zeta)=i\,\frac{\xi_y}{(1+i\zeta)^2}\, e^{-\frac{\xi^2}{1+i\zeta}},\label{linparacp}\\
&\widetilde{V}_-(\xi_x, \xi_y,\zeta)=\frac{-\xi_x}{(1+i\zeta)^2}\, e^{-\frac{\xi^2}{1+i\zeta}}. \label{linparacm}
\end{align}
\end{subequations} 
When inserted into (\ref{parpeep}) and (\ref{parpbeb}) they produce a kind of spatially varying linearly polarized Gaussian beam, if one remembers that $\varepsilon$ is tiny. The complex electric field expressed in cylindrical coordinates takes the form
\begin{subequations}\label{parlineec}
\begin{align}
\widetilde{\mathcal{E}}_{\xi_x}=&-2i\frac{\xi^2}{(1+i\zeta)^3}\,\sin 2\phi \;e^{-\frac{\xi^2}{1+i\zeta}},\label{parlinecx}\\
\widetilde{\mathcal{E}}_{\xi_y}=&\;2 i \Big[\frac{\xi^2}{(1+i\zeta)^3}\,\cos 2\phi \nonumber\\
&+\varepsilon^2\,\frac{\xi^4-2(1+i\zeta-\xi^2)^2}{(1+i\zeta)^5}\Big]\, e^{-\frac{\xi^2}{1+i\zeta}},\label{parlinecy}\\
\widetilde{\mathcal{E}}_{\zeta}=&\;4\varepsilon \xi \,\frac{2(1+i\zeta)-\xi^2}{(1+i\zeta)^4}\,\sin\phi\; e^{-\frac{\xi^2}{1+i\zeta}}.\label{parlinecz}
\end{align}
\end{subequations}
Similarly for the magnetic field one gets
\begin{subequations}\label{parlinebc}
\begin{align}
\widetilde{\mathcal{B}}_{\xi_x}=&\;-2 i \Big[\frac{\xi^2}{(1+i\zeta)^3}\,\cos 2\phi \nonumber\\
&-\varepsilon^2\,\frac{\xi^4-2(1+i\zeta-\xi^2)^2}{(1+i\zeta)^5}\Big]\, e^{-\frac{\xi^2}{1+i\zeta}},\label{parlinbcx}\\
\widetilde{\mathcal{B}}_{\xi_y}=&-2i\frac{\xi^2}{(1+i\zeta)^3}\,\sin 2\phi\; e^{-\frac{\xi^2}{1+i\zeta}},\label{parlinbcy}\\
\widetilde{\mathcal{B}}_{\zeta}=&\;4\varepsilon \xi\,\frac{2(1+i\zeta)-\xi^2}{(1+i\zeta)^4}\,\cos\phi\; e^{-\frac{\xi^2}{1+i\zeta}}.\label{parlinbcz}
\end{align}
\end{subequations}
Both transverse components of the electric (and magnetic) fields have the same coefficient ``$i$'' which means that there is no time phase shift between them, and there appears no rotation typical for circular polarization. 

It will be shown below that the associated Poynting vector exhibits a transparent backflow phenomenon. Reinstating the factor $e^{i\frac{\zeta-\tau}{2\varepsilon^2}}$, one obtains the physical fields by taking the real parts:
\begin{subequations}\label{reafie}
\begin{align}
&\bm{E}=\Re\left(e^{i\frac{\zeta-\tau}{2\varepsilon^2}}\widetilde{\bm{E}}\right),\label{reafiee}\\
&\bm{B}=\Re\left(e^{i\frac{\zeta-\tau}{2\varepsilon^2}}\widetilde{\bm{B}}\right),\label{reafieb}
\end{align}
\end{subequations}
and then the Poynting vector according to
\begin{equation}\label{poy1}
\bm{S}(\tau,\bm{\xi},\zeta)= \bm{E}\times \bm{B}.
\end{equation}
The overall constant comprising $\mu_0$, $w_0$ and $c$ is irrelevant to our further discussion and will be consequently skipped. 

\begin{figure}[h!]
\begin{center}
\includegraphics[width=0.4\textwidth,angle=0]{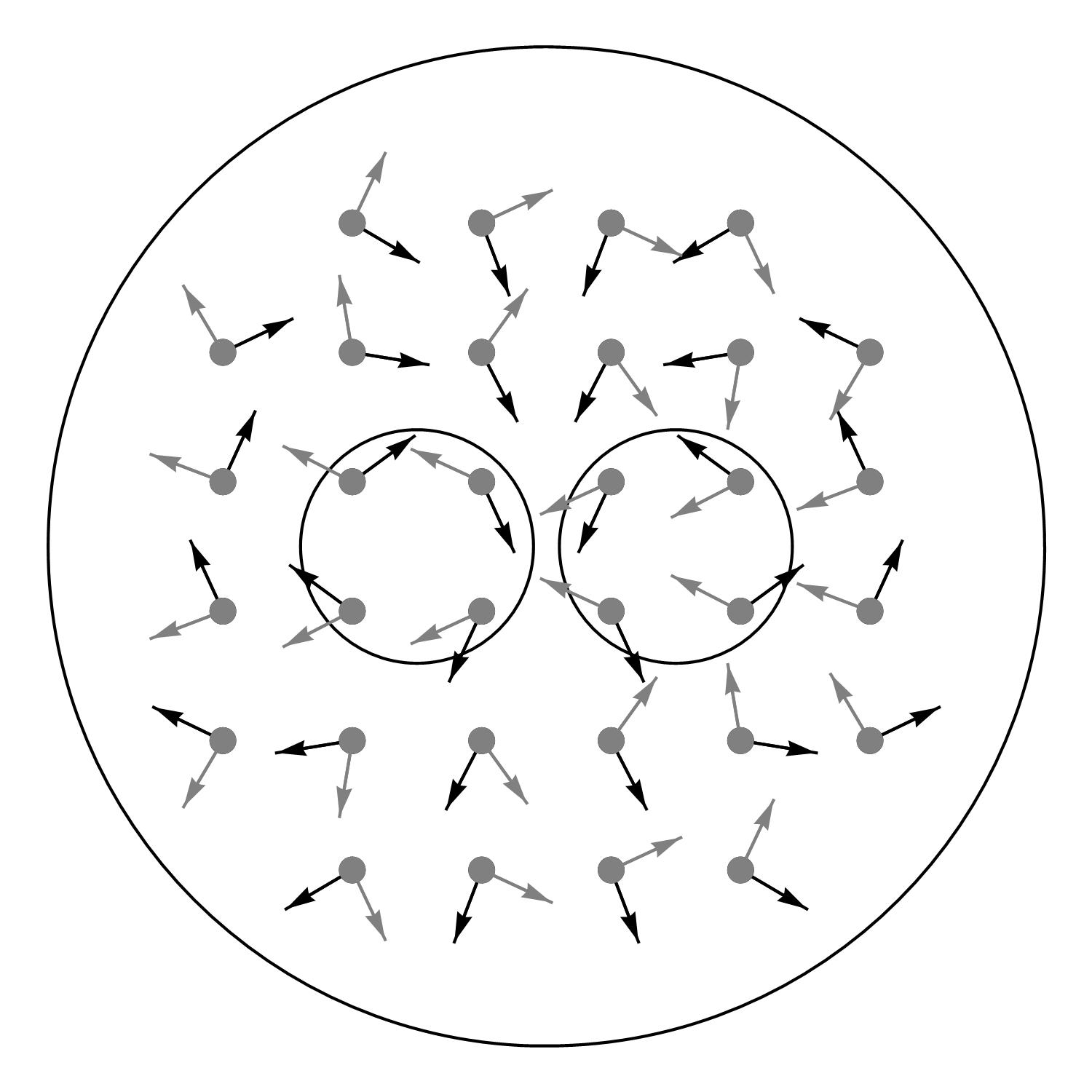}
\caption{The orientation of transverse components of electric (black arrows) and magnetic (grey arrows) fields as given by (\ref{reafie}) with (\ref{parlinecz}) and (\ref{parlinbcz}) in the plane $\zeta=0$ for a fixed value of time ($\tau=0.1$) and for $\varepsilon=0.15$. All arrows are normalized as to their length, so it does nor reflect the fields strength at a given point. The small circles indicate areas where the alternating relative orientation of both fields entails the opposite signs of the component $S_\zeta$ of the Poynting vector. The radius of the large circle equals to $0.77$.}
\label{FieldsLinC}
\end{center}
\end{figure}

In Fig.~\ref{FieldsLinC} the vectors (or rather their transverse components) $\bm{E}$ and $\bm{B}$ for certain instant of time and for $\zeta=\mathrm{const}$ are depicted. In the areas marked with small circles, the different relative orientation of the black ($\bm{E}$) and gray vectors ($\bm{B}$) can be seen, resulting in the opposite direction of the $\bm{S}$ vector. Clearly, these fields are not exactly aligned in the plane of $\zeta=\mathrm{const}$, owing to the existence of (small) longitudinal components [see (\ref{parlineec}) and (\ref{parlinebc})], but they are irrelevant for the sign of the $\zeta$ component of the Poynting vector. 

\begin{figure}[h!]
\begin{center}
\includegraphics[width=0.4\textwidth,angle=0]{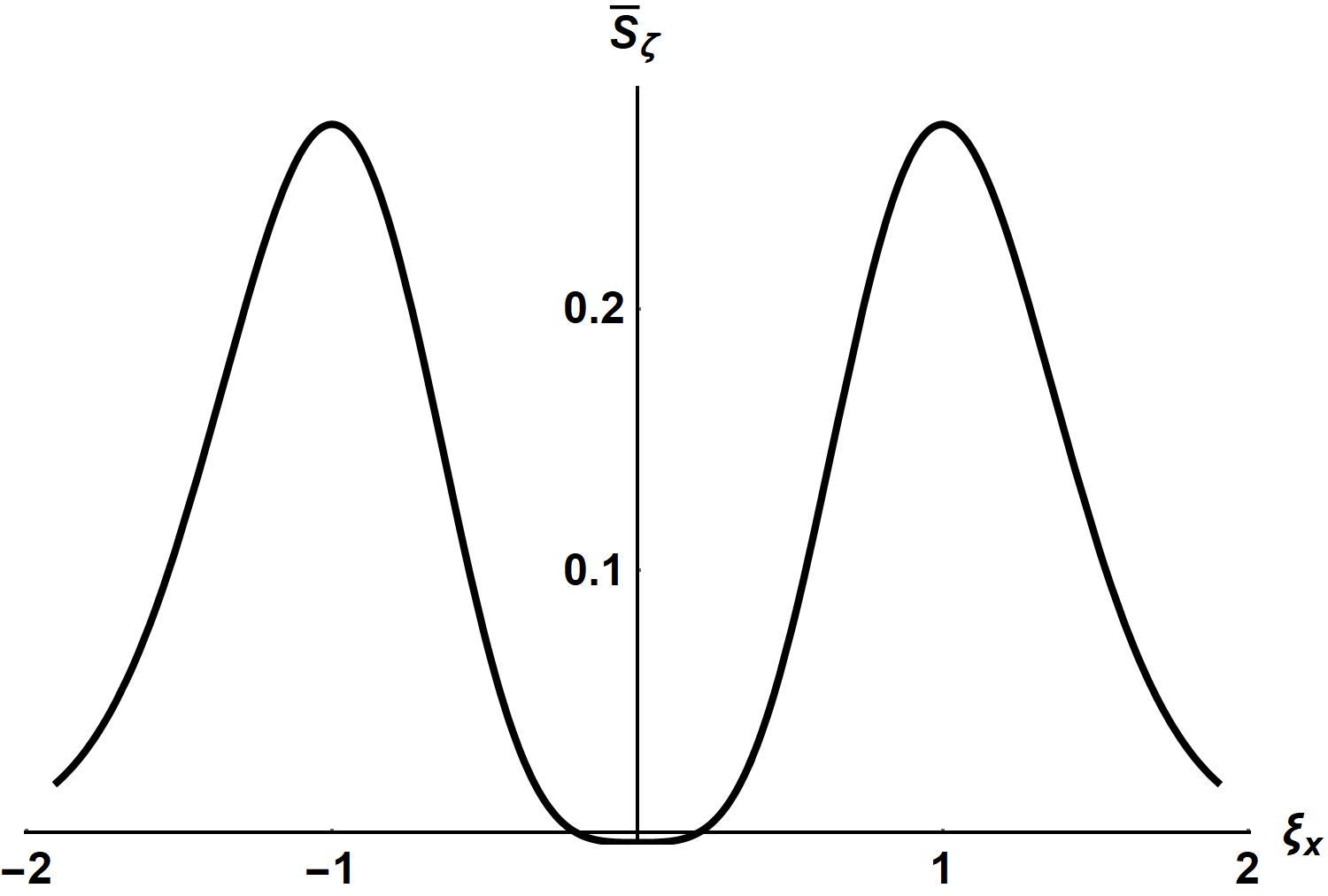}
\caption{The dependence of $\overline{S}_\zeta$ on $\xi_x$ if $\xi_y=\zeta=0$ for $\varepsilon=0.15$. Note the appearance of negative values for small $\xi_x$. }
\label{linc2d}
\end{center}
\end{figure}

What matters for the backflow phenomenon, however, is not so much the instantaneous orientation of these vectors -- although it may indicate the potential for this phenomenon to occur -- but rather $S_\zeta$ averaged over fast field oscillations, to be denoted hereafter with the symbol $\overline{S}_\zeta(\bm{\xi},\zeta)$. Leaving aside the lengthy but rather obvious transformations, we get the following result for the component in question
\begin{eqnarray}
\overline{S}_\zeta&&(\bm{\xi},\zeta)=\frac{2}{(1+\zeta^2)^3}\,e^{-\frac{2\xi^2}{1+\zeta^2}}\label{szave}\\
&&\times\Big[\xi^4-\varepsilon^4\,\frac{4\zeta^4+(2-4\xi^2+\xi^4)^2+\zeta^2(2-4\xi^2+3\xi^4)^2}{(1+\zeta^2)^2}\Big].\nonumber
\end{eqnarray}
Note the minus sign before $\varepsilon^4$ in square brackets (it is essential that the accompanying fraction is always positive). When approaching the beam axis, the first term fades out and the $\zeta$-component of the Poynting vector gets negative, entailing the emergence of the backflow. This effect is shown in Fig.~\ref{linc2d} in the form of the plot $\overline{S}_\zeta(\xi_x,\xi_y,\zeta)$ for $\zeta=0$ and $\xi_y=0$.

\begin{figure}[h!]
\begin{center}
\includegraphics[width=0.4\textwidth,angle=0]{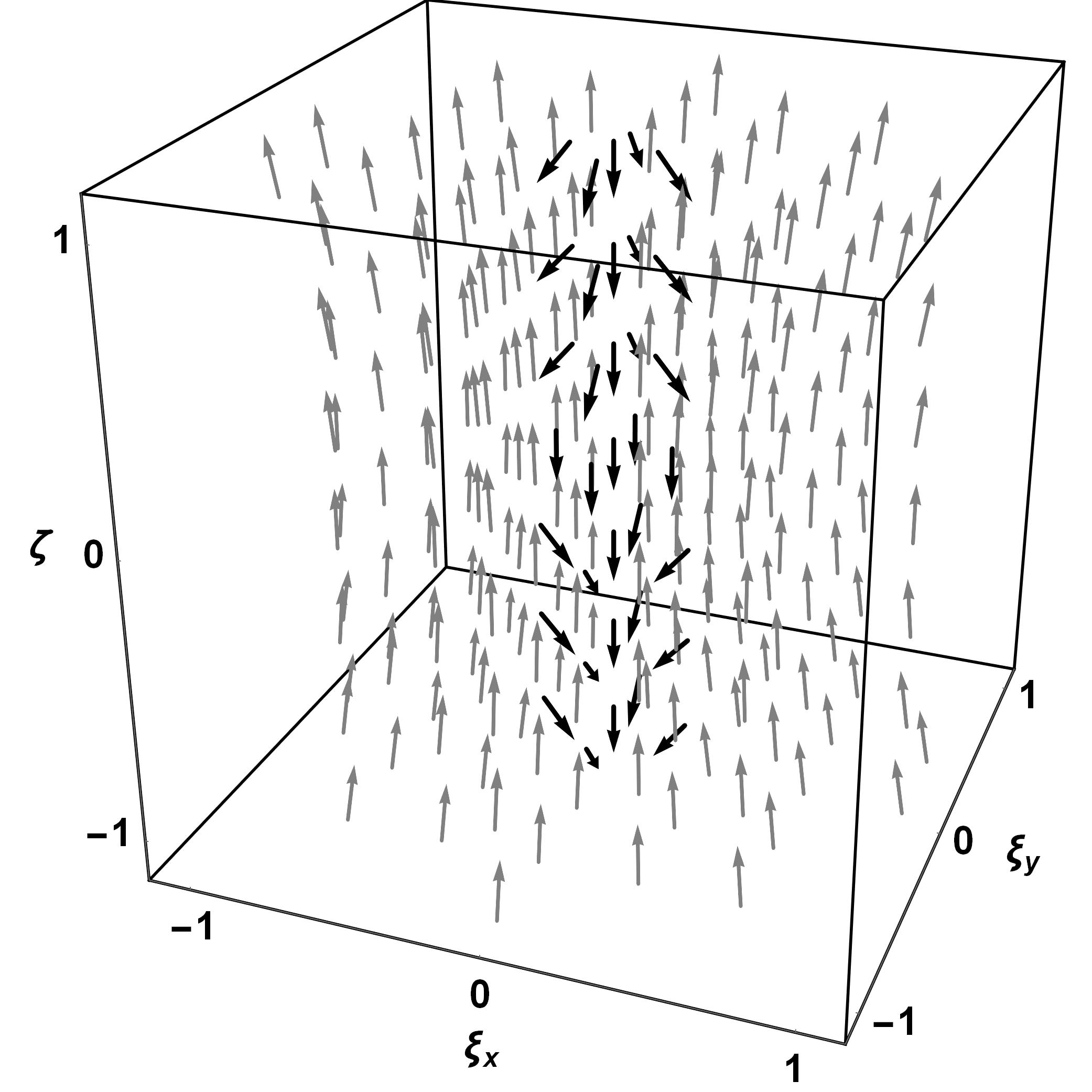}
\caption{Representation of the vector field $\overline{\bm{S}}$ for electric  and magnetic fields (\ref{parlineec}) and (\ref{parlinebc}). Thick black arrows show the retro-propagating Poynting vector along the $\zeta$-axis.}
\label{vpmlinc}
\end{center}
\end{figure}

Let us notice that the $\varepsilon$-containing expressions in the formulas  (\ref{parpeep}) and (\ref{parpbeb}) derived in \cite{trvec} entirely contribute to this phenomenon. In order to estimate the size of this contribution, we will integrate $S_\zeta$ across the beam:
\begin{equation}\label{popints}
\int d^2\xi\, \overline{S}_\zeta(\bm{\xi},\zeta)=\frac{\pi}{2}(1-3\epsilon^4).
\end{equation} 
The latter term uniquely quantifies the total amount of backflow in any transverse cross-section of the beam, owing to the fact that the two bracketed expressions in (\ref{szave}) have well-defined signs. This is an interesting and significant result: it turns out that along the entire length of the beam, the Poynting vector exhibits the steady propagation opposite to that of the beam. This is visualized in Fig. \ref{vpmlinc}.

For small values of $\varepsilon$ the second term in (\ref{popints}) is tiny, however much larger than the negative momenta content of the beam whose magnitude, as we know, is determined by the factor of $e^{-\frac{1}{2\varepsilon^2}}$. In particular, in this work, all figures are performed for the value $\varepsilon=0.15$ for which
\begin{equation}\label{warepi}
\varepsilon^4\approx 5\cdot 10^{-4},\qquad e^{-\frac{1}{2\varepsilon^2}}\approx 2\cdot 10^{-10}.
\end{equation}
Hence, the scale of the observed backflow is $6$ orders of magnitude larger than might be expected from the spectral analysis of the Gaussian beam! For smaller values of $\varepsilon$, this deviation becomes gigantic, which highlights even more the occurrence of this effect, (for example, for $\varepsilon=0.1$ it is already $18$ orders of magnitude), but it is harder to be visualized in figures.

It should be emphasized that, as seen from (\ref{popints}), the negative contribution to the integral is $\zeta$-independent: the total amount of backflow is identical in all cross sections along the beam.

\begin{figure}[h!]
\begin{center}
\includegraphics[width=0.45\textwidth,angle=0]{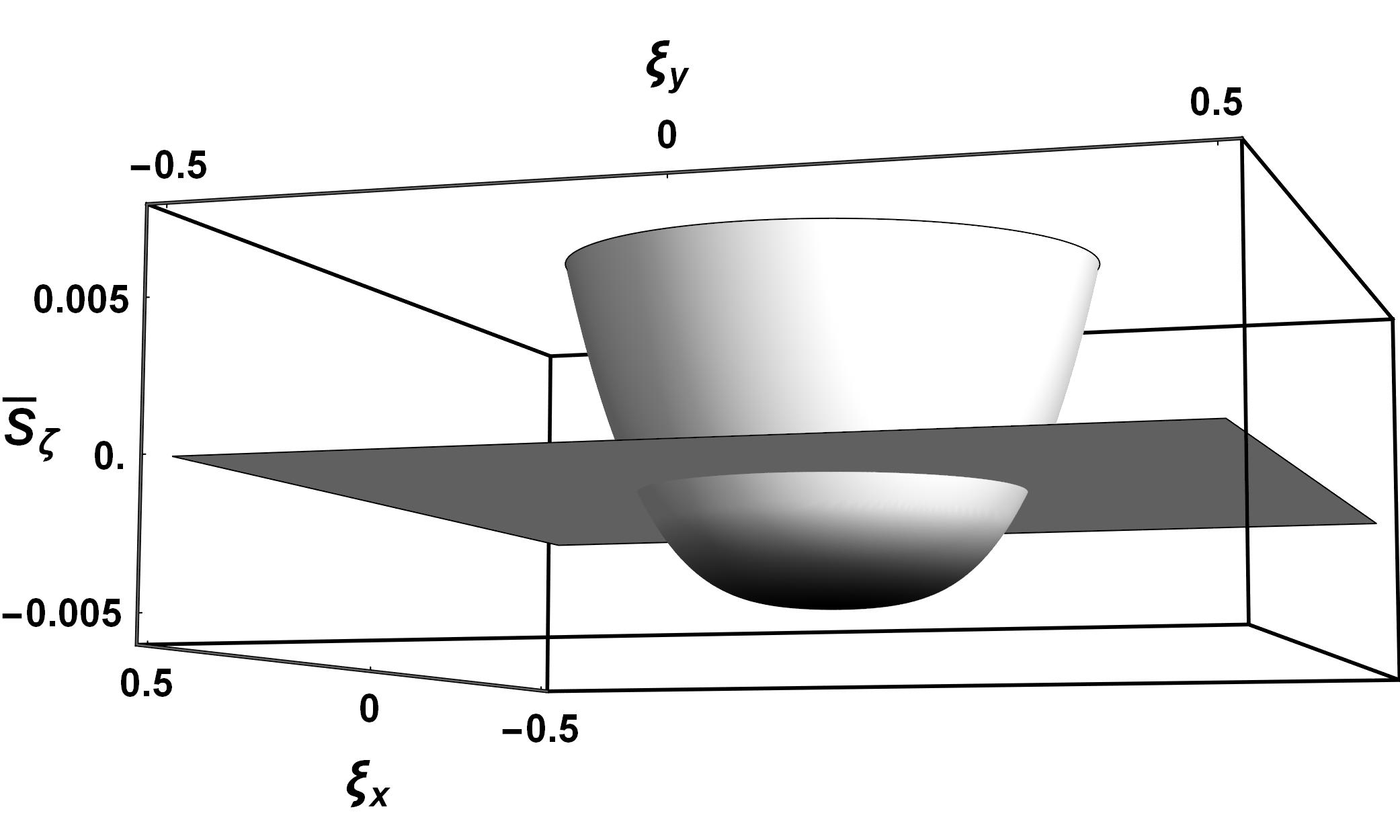}
\caption{Plot of $\overline{S}_\zeta(\xi_x,\xi_y,\zeta)$ in the plane $\zeta=0$ and for $\varepsilon=0.15$ as in Fig. \ref{linc2d}. The grey plane cuts off the negative values of $\overline{S}_\zeta$ in order to make them clearly visible.}
\label{linc3d}
\end{center}
\end{figure}

Fig.~\ref{linc3d} displays the dependence $\overline{S}_\zeta(\bm{\xi},\zeta)$ in the surface $\zeta=0$ in three-dimensional form. The auxiliary plane drawn at zero value exposes the area where $\overline{S}_\zeta$ is negative and the backflow appears.

\begin{figure}[h!]
\begin{center}
\includegraphics[width=0.4\textwidth,angle=0]{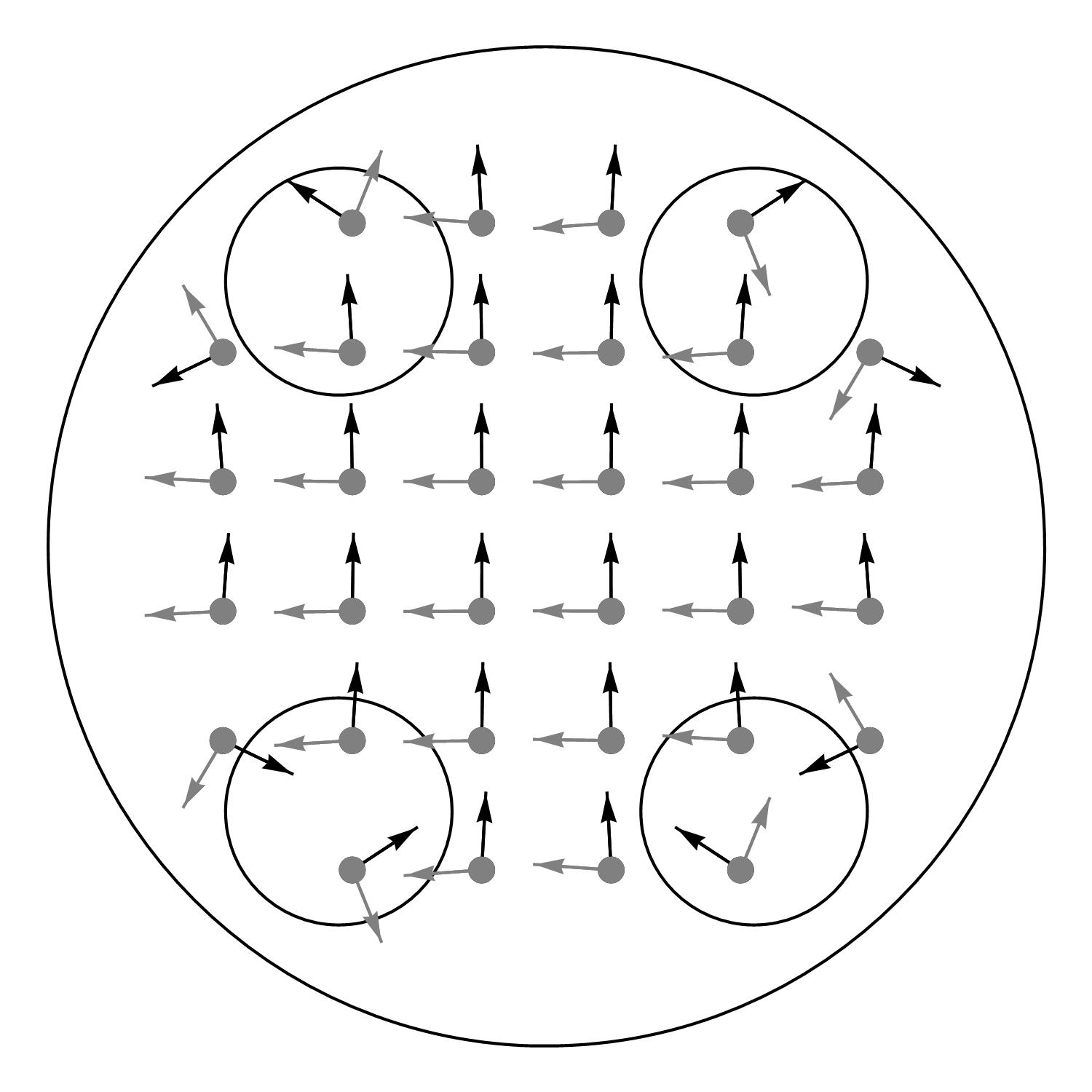}
\caption{Same as Fig. \ref{FieldsLinC}, but for fields defined by ``potentials'' (\ref{lipara}). The radius of the large circle equals $1.32$.}
\label{FieldsLinA}
\end{center}
\end{figure}

The situation is somewhat different below, where it will be shown that by choosing linear polarization of the Gaussian beam \cite{trvec} with almost constant alignment of the fields across the beam, the backflow region can be made annular. Let us for example set
\begin{subequations}\label{lipara}
\begin{align}
&\widetilde{V}_+(\xi, \phi,\zeta)=i\,\frac{\xi_y}{(1+i\zeta)^2}\, e^{-\frac{\xi^2}{1+i\zeta}},\label{linparap}\\
&\widetilde{V}_-(\xi, \phi,\zeta)=\frac{\xi_x}{(1+i\zeta)^2}\, e^{-\frac{\xi^2}{1+i\zeta}}. \label{linparam}
\end{align}
\end{subequations} 
The expressions for the complex electric field are then as follows
\begin{subequations}\label{palineec}
\begin{align}
\widetilde{\mathcal{E}}_{\xi_x}=&2i\varepsilon^2\xi^2\,\frac{3(1+i\zeta)-\xi^2}{(1+i\zeta)^5}\, \sin 2\phi\;e^{-\frac{\xi^2}{1+i\zeta}},\label{palinecx}\\
\widetilde{\mathcal{E}}_{\xi_y}=&\;2 i \Big[\frac{1+i\zeta-\xi^2}{(1+i\zeta)^3} \nonumber\\
&-\varepsilon^2\xi^2\,\frac{3(1+i\zeta)-\xi^2}{(1+i\zeta)^5}\Big]\,\cos 2\phi\; e^{-\frac{\xi^2}{1+i\zeta}},\label{palinecy}\\
\widetilde{\mathcal{E}}_{\zeta}=&\;4\varepsilon \xi\,\frac{2(1+i\zeta)-\xi^2}{(1+i\zeta)^4}\,\sin\phi \; e^{-\frac{\xi^2}{1+i\zeta}}.\label{palinecz}
\end{align}
\end{subequations}
and for the magnetic field
\begin{subequations}\label{palinebc}
\begin{align}
\widetilde{\mathcal{B}}_{\xi_x}=&\;-2 i \Big[\frac{1+i\zeta-\xi^2}{(1+i\zeta)^3} \nonumber\\
&+\varepsilon^2\xi^2\,\frac{3(1+i\zeta)-\xi^2}{(1+i\zeta)^5}\,\cos 2\phi\Big]\, e^{-\frac{\xi^2}{1+i\zeta}},\label{palinbcx}\\
\widetilde{\mathcal{B}}_{\xi_y}=&-2i\varepsilon^2\xi^2\,\frac{3(1+i\zeta)-\xi^2}{(1+i\zeta)^3}\,\sin 2\phi\; e^{-\frac{\xi^2}{1+i\zeta}},\label{palinbcy}\\
\widetilde{\mathcal{B}}_{\zeta}=&\;-4\varepsilon \xi\,\frac{2(1+i\zeta)-\xi^2}{(1+i\zeta)^4}\, \cos\phi\; e^{-\frac{\xi^2}{1+i\zeta}}.\label{palinbcz}
\end{align}
\end{subequations}
Note that for $\varepsilon\ll 1$ the electric field gets actually aligned along the $\xi_y$ axis and magnetic field along the $\xi_x$ axis. This is, however, changed in the vicinity of the circle $\xi=1$, especially for small values of $\zeta$, where the second term in the expressions for $\widetilde{\mathcal{E}}_{\xi_y}$ and $\widetilde{\mathcal{B}}_{\xi_x}$ may become predominant. An important point is the difference in signs at $\varepsilon^2$: since these signs are distinct, the sign of $\overline{S}_\zeta$ may also change, so this is the region where backflow is likely to occur. This effect is demonstrated in Fig.~\ref{FieldsLinA}.

\begin{figure}[h!]
\begin{center}
\includegraphics[width=0.4\textwidth,angle=0]{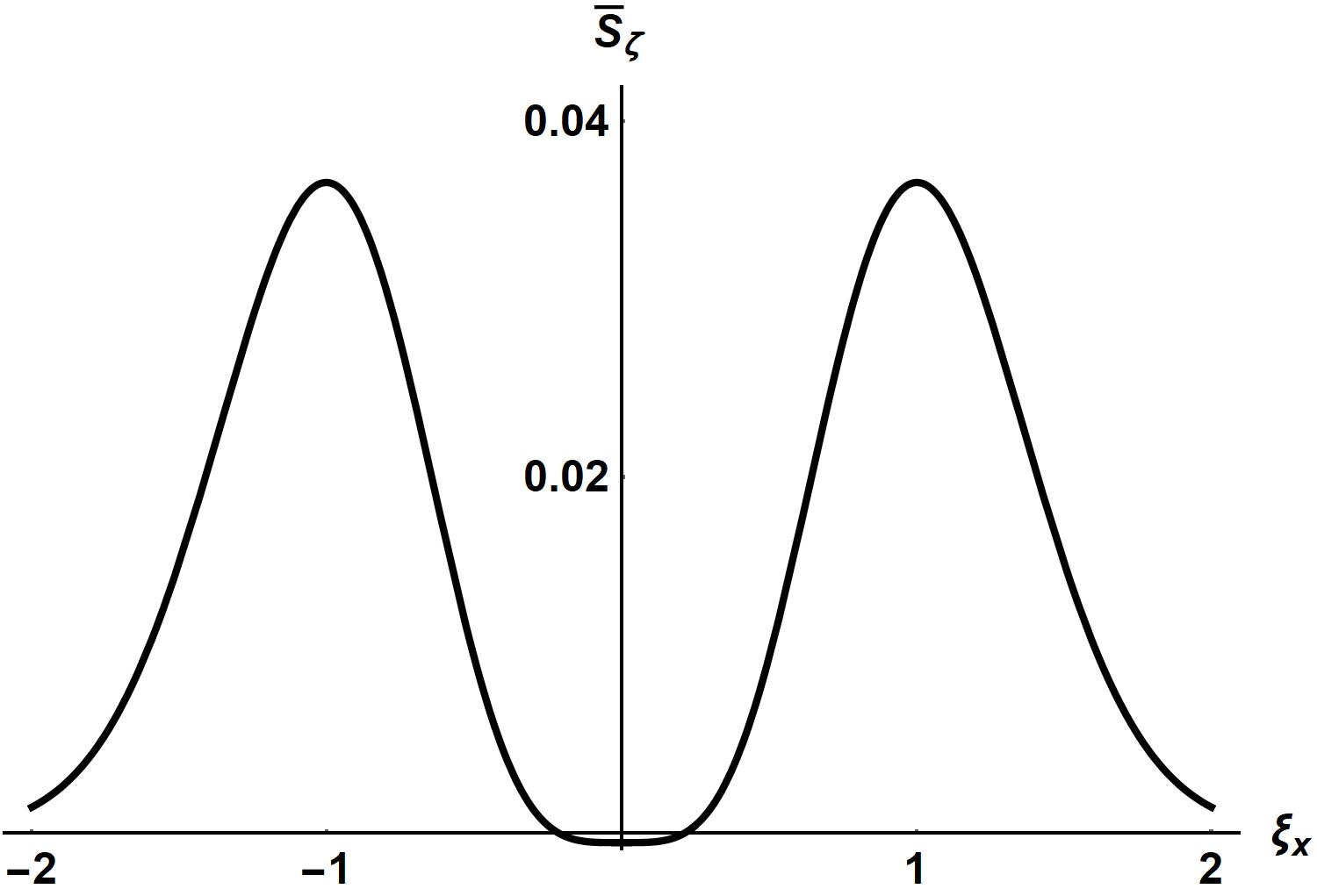}
\caption{Same as Fig.~\ref{linc2d} but for but for fields defined by ``potentials'' (\ref{lipara}) and for $\xi_y=1$.}
\label{lina2d}
\end{center}
\end{figure}

\begin{figure}[h!]
\begin{center}
\includegraphics[width=0.45\textwidth,angle=0]{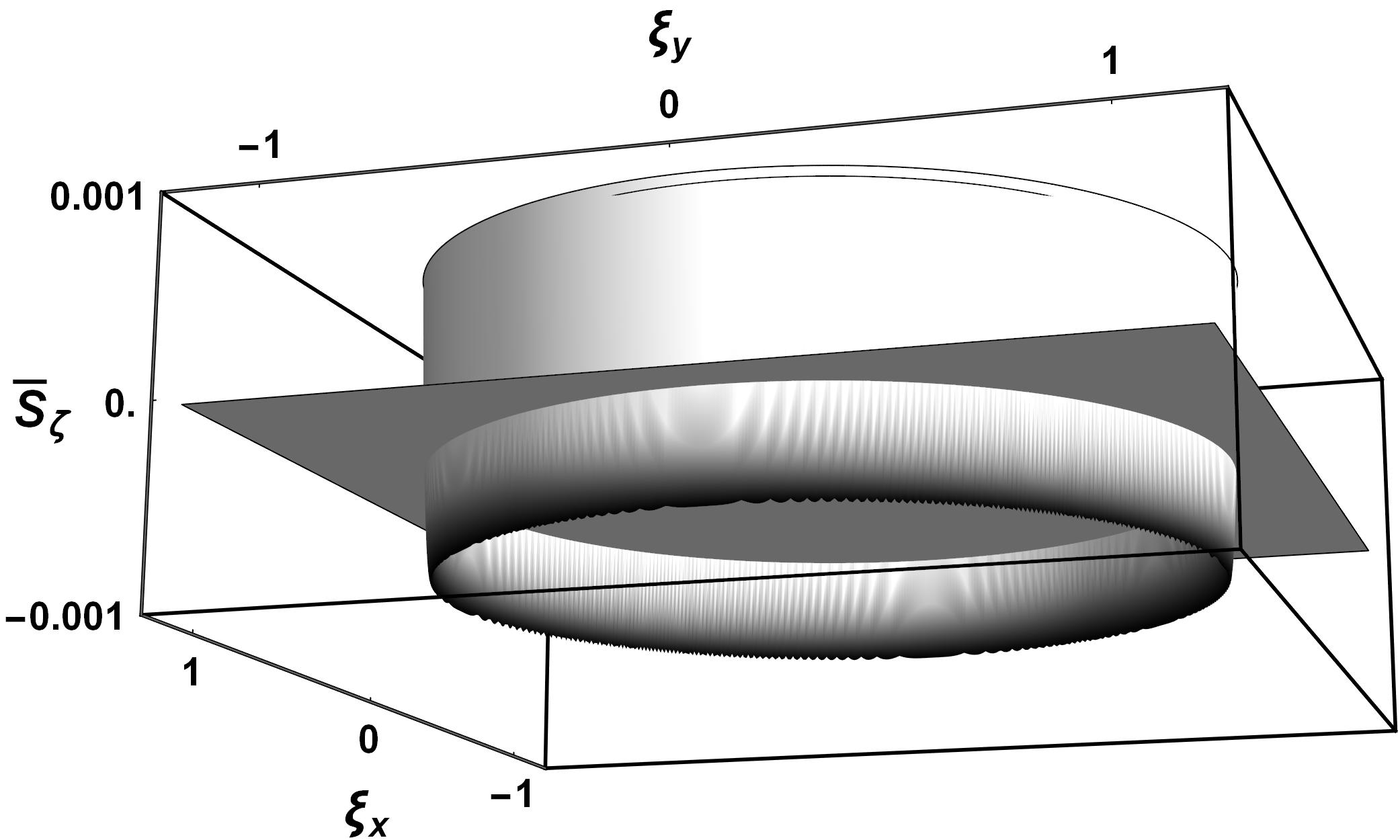}
\caption{Same as Fig.~\ref{linc3d} but for but for fields defined by ``potentials'' (\ref{lipara}). The negative values of $\overline{S}_\zeta$ are arranged around the ring $\xi=1$.}
\label{lina3d}
\end{center}
\end{figure}

The third component on the Poynting vector averaged with respect to time is now given by
\begin{eqnarray}
\overline{S}_\zeta(\bm{\xi},\zeta)&=&\frac{2}{(1+\zeta^2)^3}\,e^{-\frac{2\xi^2}{1+\zeta^2}}\label{szavea}\\
&&\times\Big[(1-\xi^2)^2+\zeta^2-\varepsilon^4\xi^4\,\frac{(3-\xi^2)^2+9\zeta^2}{(1+\zeta^2)^2}\Big],\nonumber
\end{eqnarray}
clearly showing that it becomes negative for $\xi\approx 1$ and close to the focal plane of the beam.
That effect can be seen in the Figs~\ref{lina2d} and~\ref{lina3d}. Although the former one is apparently similar to Fig.~\ref{linc2d}, it should be noted that the plot is now performed for $\xi_y=1$ so that the negative values are located on the ring. This is explicitly demonstrated in Fig.~\ref{lina3d} in the form of the three-dimensional graph.

Integrating $\overline{S}_\zeta(\bm{\xi},\zeta)$ over the cross section of the beam, identically as before, again expression (\ref{popints}) is obtained. Since the coefficient function of $\varepsilon^4$ in (\ref{szavea}) is negative for any $\zeta$, it is clear once again that $\varepsilon$ contributions in (\ref{palineec}) and (\ref{palinebc}) are entirely responsible for the appearance of the backflow.

Similar structures of retrograde flow as demonstrated in Figs \ref{linc3d} and \ref{lina3d} were identified in \cite{kotkot}.

The question arises whether the observed effects can be noticed for other polarizations, such as (almost) radial or azimuthal. In the former case one chooses (they in fact belong to the order $n=0$ Gaussian beams):
\begin{subequations}\label{radpara}
\begin{align}
&\widetilde{V}_+(\xi, \phi,\zeta)=\frac{1}{1+i\zeta}\, e^{-\frac{\xi^2}{1+i\zeta}},\label{lradparap}\\
&\widetilde{V}_-(\xi, \phi,\zeta)=0. \label{radparam}
\end{align}
\end{subequations} 
and for the latter one:
\begin{subequations}\label{asipara}
\begin{align}
&\widetilde{V}_+(\xi, \phi,\zeta)=0,\label{lasiparap}\\
&\widetilde{V}_-(\xi, \phi,\zeta)=\frac{i}{1+i\zeta}\, e^{-\frac{\xi^2}{1+i\zeta}}. \label{asiparam}
\end{align}
\end{subequations}
The straightforward calculation similar to the above leads to the conclusion that in both cases the third component of $\overline{\bm{S}}$ is identical:
\begin{equation}\label{radazy}
\overline{S}_\zeta(\bm{\xi},\zeta)=\frac{2\xi^2}{(1+\zeta^2)^2}\,e^{-\frac{2\xi^2}{1+\zeta^2}}\Big[1-\varepsilon^4\,\frac{4\zeta^2+(2-\xi^2)^2}{(1+\zeta^2)^2}\Big].
\end{equation} 
However, this expression turns negative only at very large values of $\xi$ (of order of $\frac{1}{\varepsilon}$), for which the beam is strongly suppressed due to the Gaussian damping factor. The paraxial approximation is obviously violated there. Therefore, from the practical point of view, for these polarizations the backflow phenomenon cannot be observed even though the cross-sectional integral gives comparable result to the former:
\begin{equation}\label{popintsa}
\int d^2\xi\, \overline{S}_\zeta(\bm{\xi},\zeta)=\frac{\pi}{4}(2-3\epsilon^4).
\end{equation}

\subsection{The $n=2$ case}\label{endwa}

\begin{figure}[h!]
\begin{center}
\includegraphics[width=0.4\textwidth,angle=0]{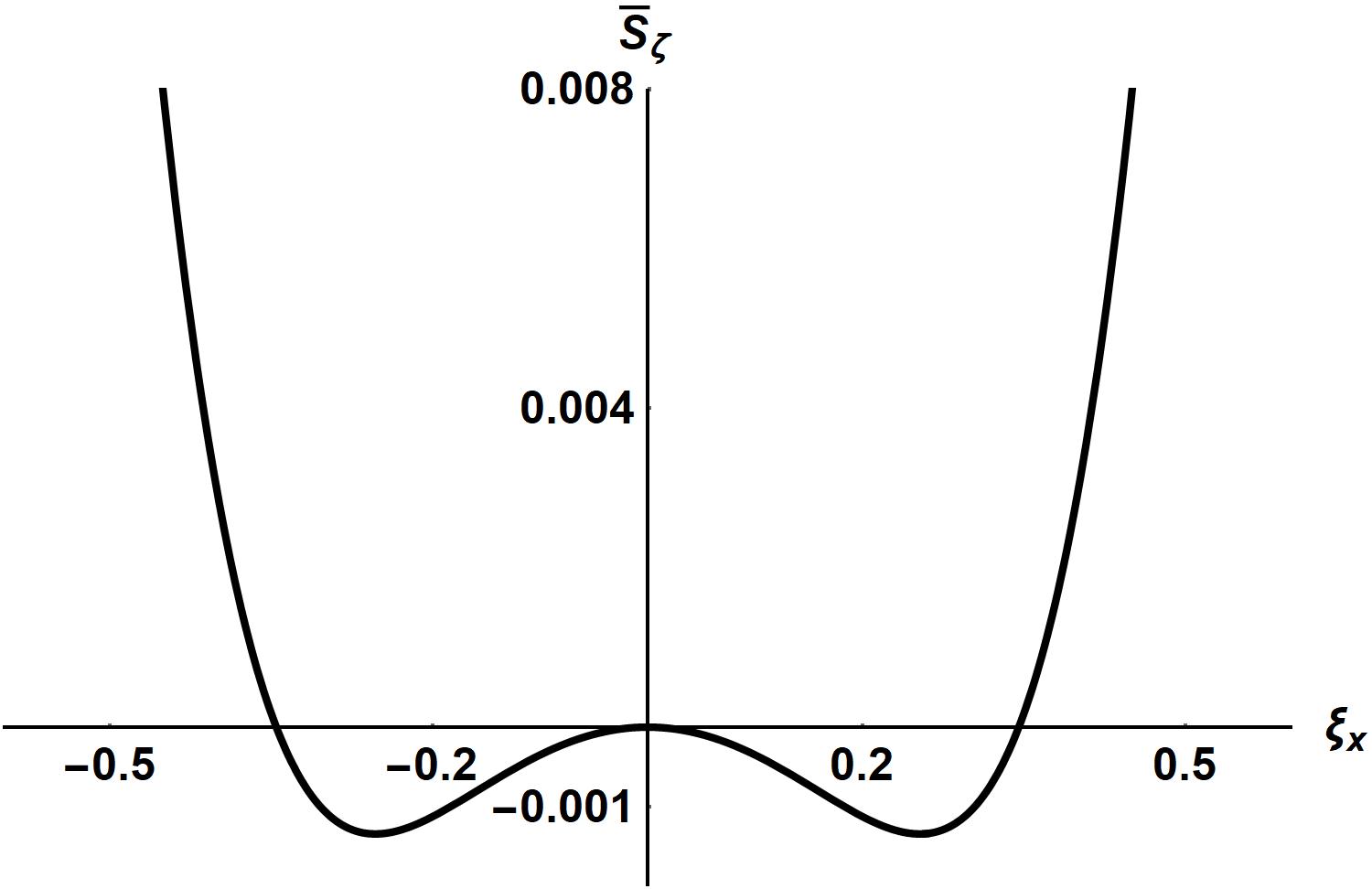}
\caption{Same as Fig.~\ref{linc2d} but for but for fields obtained from potentials (\ref{drb}).}
\label{kwb2d}
\end{center}
\end{figure}

\begin{figure}[h!]
\begin{center}
\includegraphics[width=0.45\textwidth,angle=0]{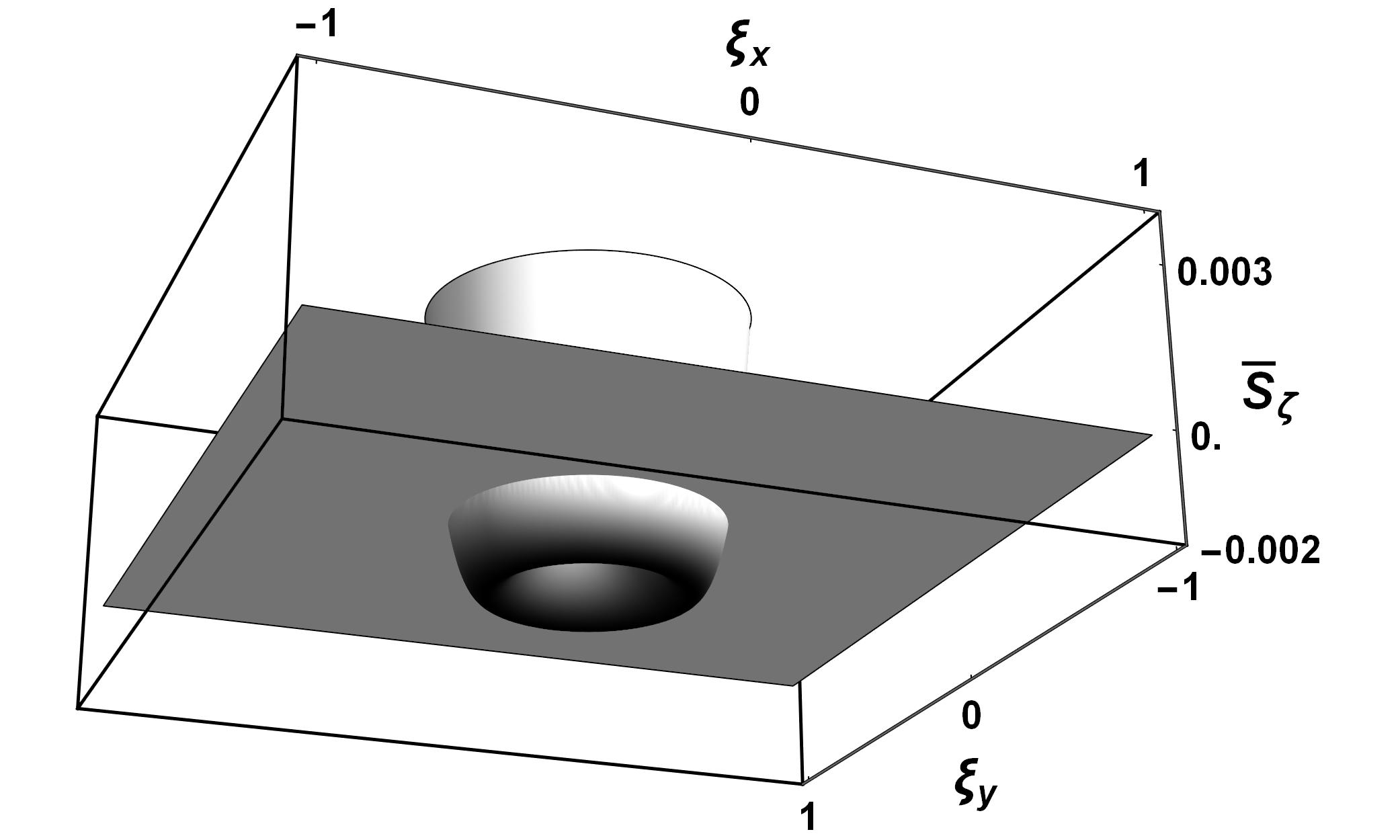}
\caption{Same as Fig.~\ref{linc3d} but for fields defined by ``potentials'' (\ref{drb}). The negative values $\overline{S}_\zeta$ are arranged around the focus.}
\label{kwb3d}
\end{center}
\end{figure}

In this section we will present a couple of examples of the $n=2$ beams exhibiting local backflow phenomenon. First we choose the potentials in the following form
\begin{subequations}\label{drb}
\begin{align}
&\widetilde{V}_+(\xi_x, \xi_y,\zeta)=\frac{\xi_x^2-\xi_y^2}{(1+i\zeta)^3}\, e^{-\frac{\xi^2}{1+i\zeta}},\label{drbp}\\
&\widetilde{V}_-(\xi_x, \xi_y,\zeta)=\frac{-2i\xi_x\xi_y}{(1+i\zeta)^3}\, e^{-\frac{\xi^2}{1+i\zeta}}. \label{drbm}
\end{align}
\end{subequations} 
In what follows, in order to avoid unnecessary lengthening of the work, the formulas for the fields $\widetilde{\bm{\mathcal{E}}}$ and $\widetilde{\bm{\mathcal{B}}}$ obtained upon inserting $\widetilde{V}_\pm$ into (\ref{parpeep}) and (\ref{parpbeb}) will no longer be written out explicitly. We will limit ourselves to the third component of the Poynting vector which is essential. Using (\ref{drb}) it can be derived in a straightforward way that
\begin{eqnarray}
\overline{S}_\zeta(\bm{\xi},\zeta)&=&\frac{2\xi^2}{(1+\zeta^2)^4}\,e^{-\frac{2\xi^2}{1+\zeta^2}}\Big[\xi^4-\varepsilon^4\times\label{szavekw}\\
&&\frac{36\zeta^4+24 \zeta^2 (3 - 3 \xi^2 + \xi^4)+(6 - 6 \xi^2 + \xi^4)^2}{(1+\zeta^2)^2}\Big].\nonumber
\end{eqnarray}
It is simple to verify that the fraction accompanying $\varepsilon^4$ is always positive, so the backflow phenomenon can appear in this configuration when $\xi^4$ is sufficiently small. That this is indeed the case can be explicitly seen by considering the sign of $\overline{S}_\zeta$ for instance in the plane $\zeta=0$: 
\begin{equation}\label{dlaze}
\overline{S}_\zeta(\bm{\xi},\zeta)=2\xi^2\,e^{-2\xi^2}\Big[\xi^4-\varepsilon^4(6 - 6 \xi^2 + \xi^4)^2\Big].
\end{equation}
It is clear that this expression becomes negative near the propagation axis of the wave, although on the axis itself  $\overline{S}_\zeta$ vanishes owing to the factor $\xi^2$. The function $\overline{S}_\zeta(\xi_x,0,0)$ is plotted in Fig.~\ref{kwb2d}.
The graph representing the full function $\overline{S}_\zeta(\bm{\xi},0)$ is depicted in the three-dimensional figure~\ref{kwb3d}. The region of backflow is clearly visible.

Just as before, the total backflow in any section is identical, which is obvious once the following integral is calculated:
\begin{equation}\label{popintskwb}
\int d^2\xi\, \overline{S}_\zeta(\bm{\xi},\zeta)=\frac{3\pi}{4}(1-5\epsilon^4).
\end{equation} 
This is a straightforward consequence of the structure of the expression (\ref{szavekw}) [and similarly (\ref{szave}) or (\ref{szavea})] -- the sign of the $\varepsilon^4$ term does not depend on $\zeta$.

\begin{figure}[h!]
\begin{center}
\includegraphics[width=0.4\textwidth,angle=0]{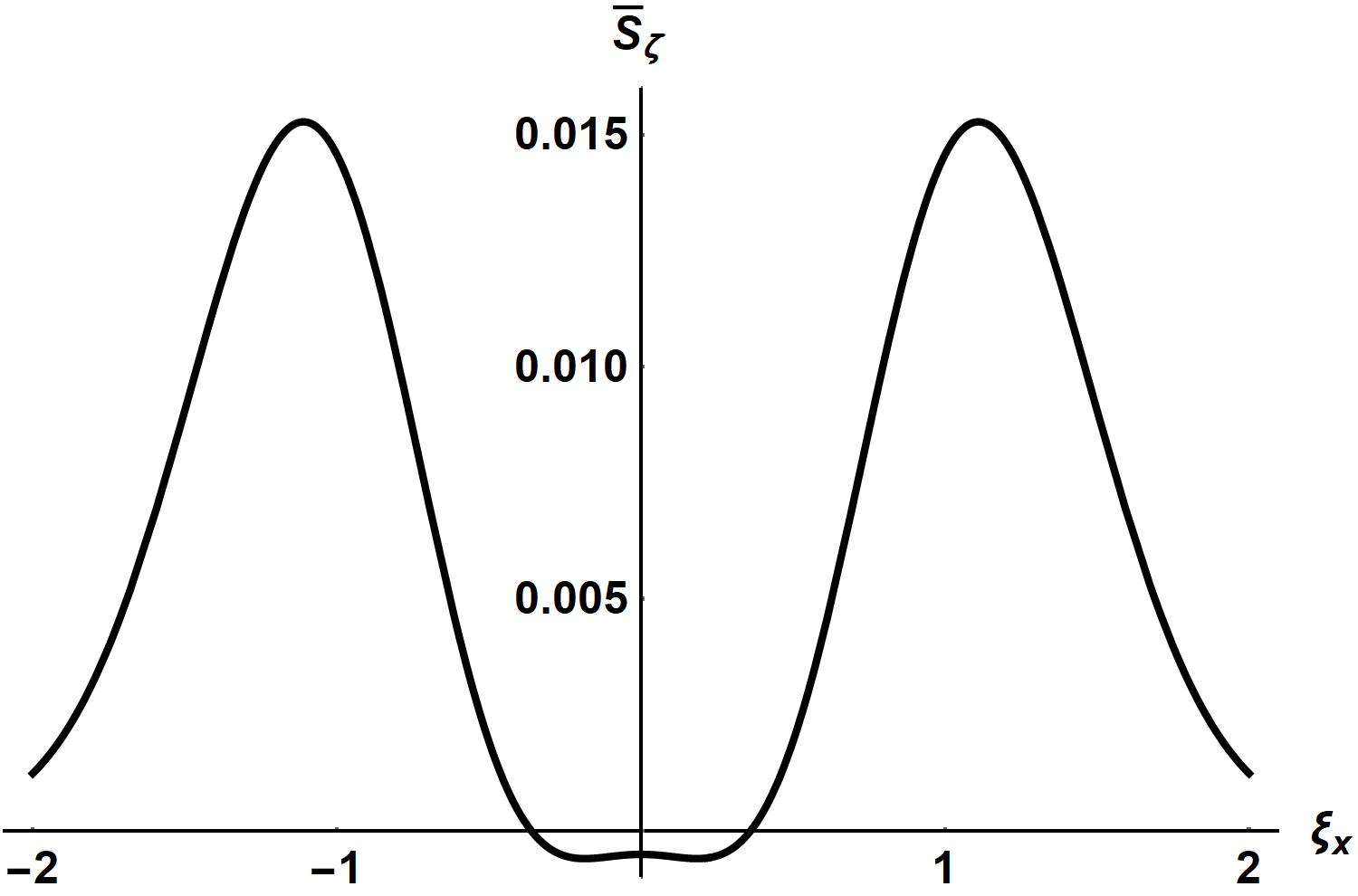}
\caption{Same as Fig.~\ref{linc2d} but for but for fields obtained from potentials (\ref{draa}) and for $\xi_y=1.4$.}
\label{kwa2d}
\end{center}
\end{figure}

\begin{figure}[h!]
\begin{center}
\includegraphics[width=0.45\textwidth,angle=0]{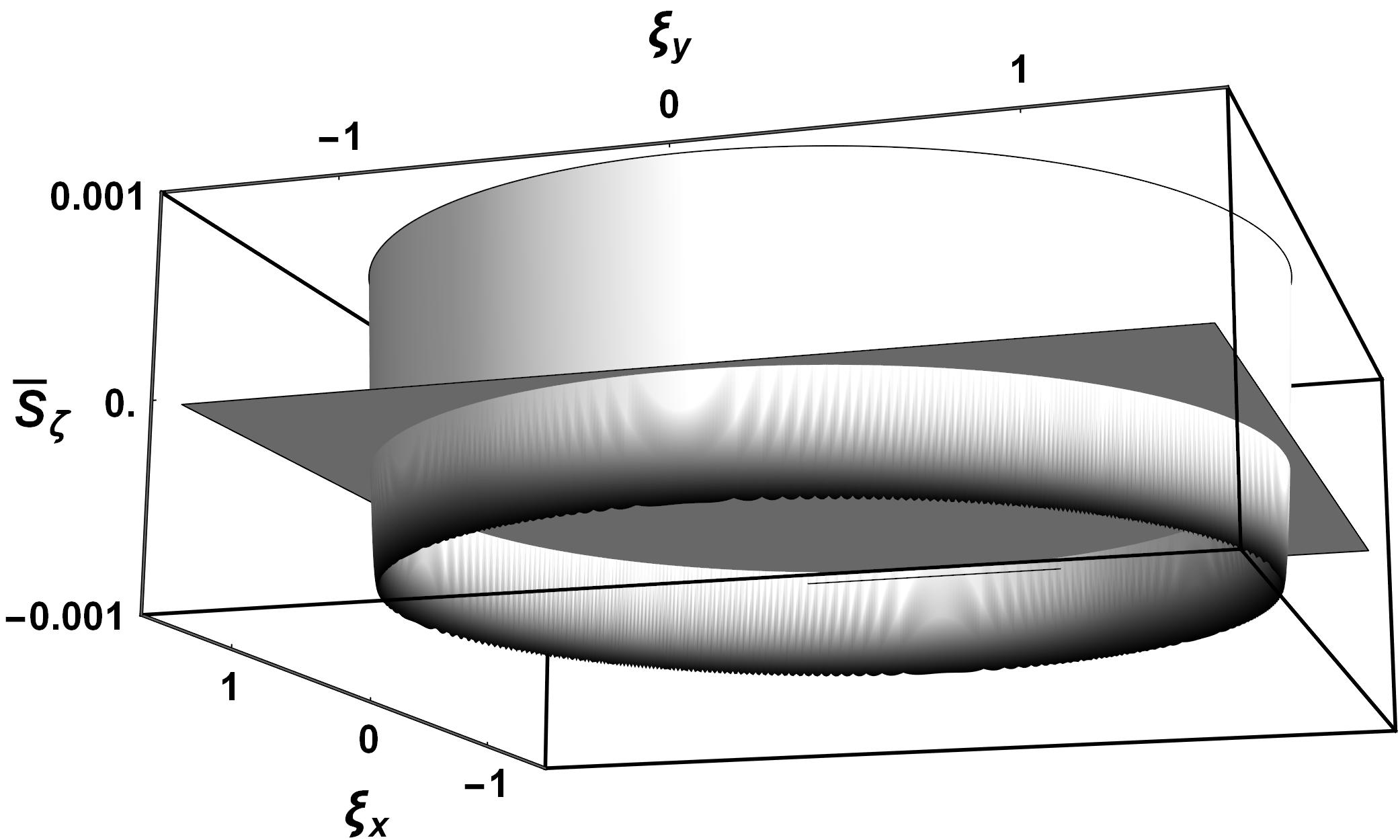}
\caption{Same as Fig.~\ref{linc3d} but for but for fields defined by potentials (\ref{draa}). The negative values of $\overline{S}_\zeta$ are arranged around the ring $\xi=\sqrt{2}$.}
\label{kwa3d}
\end{center}
\end{figure}

An annular backflow area, similar to that of Fig.~\ref{lina3d} for $n=1$, can now be obtained by simply changing the sign of $\widetilde{V}_-$:
\begin{subequations}\label{draa}
\begin{align}
&\widetilde{V}_+(\xi_x, \xi_y,\zeta)=\frac{\xi_x^2-\xi_y^2}{(1+i\zeta)^3}\, e^{-\frac{\xi^2}{1+i\zeta}},\label{draap}\\
&\widetilde{V}_-(\xi_x, \xi_y,\zeta)=\frac{2i\xi_x\xi_y}{(1+i\zeta)^3}\, e^{-\frac{\xi^2}{1+i\zeta}}. \label{draam}
\end{align}
\end{subequations} 
With this choice one gets
\begin{eqnarray}
\overline{S}_\zeta(\bm{\xi},\zeta)&=&\frac{2\xi^2}{(1+\zeta^2)^4}\,e^{-\frac{2\xi^2}{1+\zeta^2}}\Big[(2-\xi^2)^2+4\zeta^2\nonumber\\
&&-\varepsilon^4\xi^4\,\frac{(4-\xi^2)^2+16\zeta^2}{(1+\zeta^2)^2}\Big].\label{szavekwa}
\end{eqnarray}
and in the focal plane
\begin{equation}\label{dlazea}
\overline{S}_\zeta(\bm{\xi},\zeta)=2\xi^2\,e^{-2\xi^2}\Big[(2-\xi^2)^2-\varepsilon^4\xi^4(4-\xi^2)^2\Big].
\end{equation}
As it is visible, the ring of maximal backflow is roughly located at $\xi=\sqrt{2}$ as seen in Figs~\ref{kwa2d} and \ref{kwa3d} (one should note that the former plot is performed for $\xi_y=1.4$).

The total amount of backflow in any cross-section can, as usual, be read from the integral
\begin{equation}\label{popintskwa}
\int d^2\xi\, \overline{S}_\zeta(\bm{\xi},\zeta)=\frac{3\pi}{4}(1-5\epsilon^4).
\end{equation} 

\begin{figure}[h!]
\begin{center}
\includegraphics[width=0.45\textwidth,angle=0]{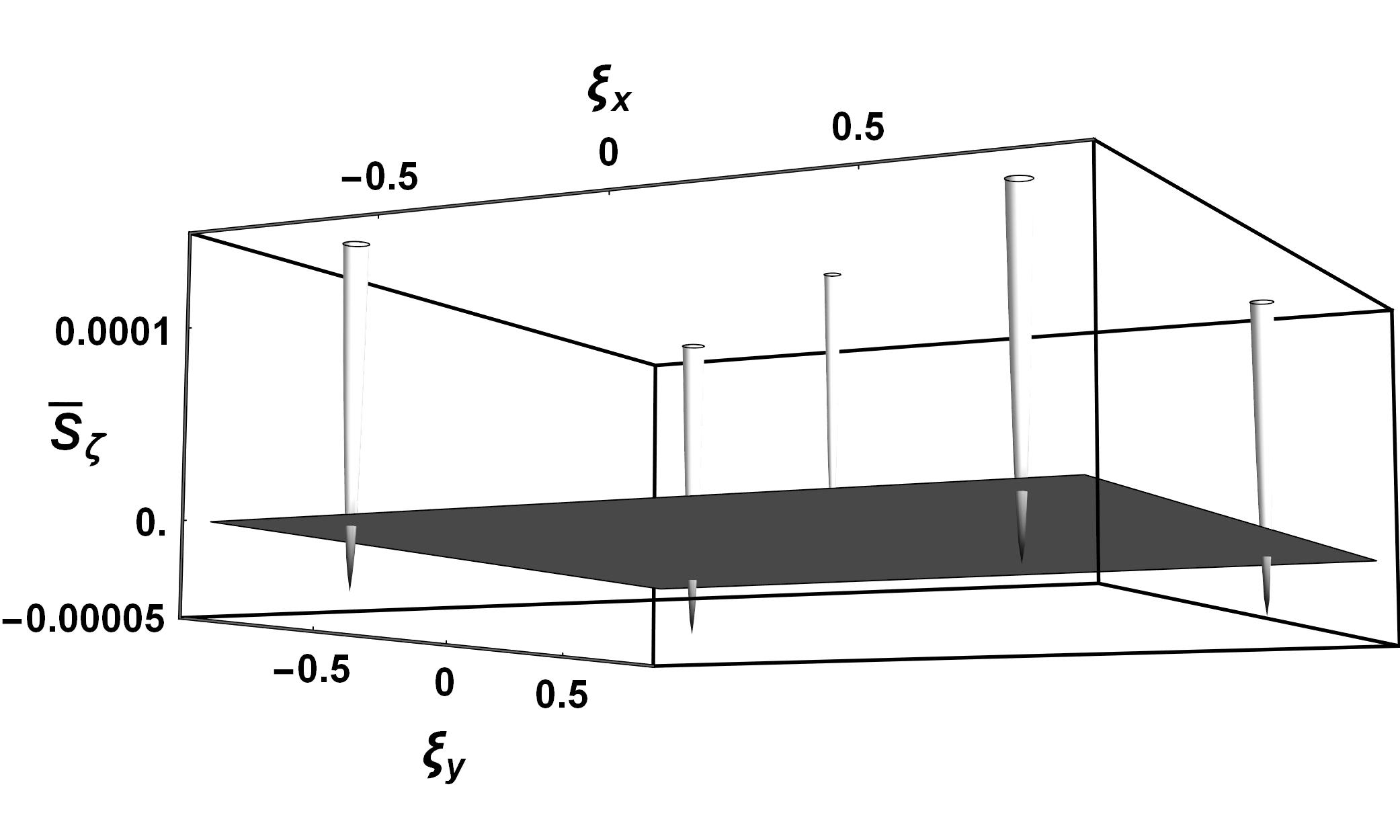}
\caption{Same as Fig.~\ref{linc3d} but for but for fields defined by potentials (\ref{drba}).}
\label{kwc3d}
\end{center}
\end{figure}

At the end we would like to add that the backflow areas can have entirely different character, as shown in Fig. \ref{kwc3d}
for the following potentials:
\begin{subequations}\label{drba}
\begin{align}
&\widetilde{V}_+(\xi_x, \xi_y,\zeta)=\frac{\xi_x\xi_y}{(1+i\zeta)^3}\, e^{-\frac{\xi^2}{1+i\zeta}},\label{drbap}\\
&\widetilde{V}_-(\xi_x, \xi_y,\zeta)=0. \label{drbam}
\end{align}
\end{subequations} 
To avoid lengthy expression the relevant component of $\overline{\bm{S}}$ will be written out only in the focal plane:
\begin{eqnarray}
\overline{S}_\zeta(\xi,\phi,0)&=&\frac{1}{4}\,\xi^2e^{-2\xi^2}\Big\{2-\xi^2(2-\xi^2)(1-\cos 4\phi)\nonumber\\
&&-\varepsilon^4\big[18-36\xi^2+32\xi^4-10\xi^6+\xi^8\label{szavekwas}\\
&&+\xi^2(24-30\xi^2+10\xi^4-\xi^6) \cos 4\phi\big]\Big\}.\nonumber
\end{eqnarray}
It is simple to show that for $\cos 4\phi =-1$ and $\xi=1$ the first two terms vanish and the behavior of the vector $\overline{\bm{S}}$ is determined by those accompanied by $\varepsilon^4$ which in total amount to negative value of $-\frac{1}{2e^2}\,\varepsilon ^4$.

The integration over the beam cross section for any $\zeta$ gives
\begin{equation}\label{popintskwas}
\int d^2\xi\, \overline{S}_\zeta(\bm{\xi},\zeta)=\frac{3\pi}{32}(1-5\epsilon^4).
\end{equation}

\section{Summary and conclusions}

The exact solutions of the Maxwell paraxial equations include terms proportional to $\varepsilon=\frac{\lambda}{2\pi w_0}$. Vector Gaussian beams derived from these equations exhibit backflow in the Poynting vector depending, however, on the polarization state. For selected beams, the backflow regions are located either on the axis of wave propagation or have an annular character. For some polarizations like radial or azimuthal, the backflow disappears. 

Depending on the instant of time, areas can be seen in which the mutual orientation of the electric and magnetic field vectors is reversed compared to adjacent domains, leading to different signs of the $z$-component of the vector $\bm{E}\times\bm{B}$. What should be emphasized, however, is that the appearing retrograde flow  is stationary and is maintained in the energy flux after time averaging of the $z$-component of the Poynting vector.  

The magnitude of this flow is determined by the parameter $\varepsilon^4$. For typical (i.e., small) values of $\varepsilon$, the magnitude of the backflow is many orders of magnitude larger than the exponentially small content of negative momenta in the Gaussian beam, which is proportional to $e^{-\frac{1}{2\varepsilon^2}}$.

The universal, i.e., two scalar-potential-dependent solutions of the Maxwell paraxial equations can be used in an analogous way to study backflow in other beams. The retro-propagation for typical laser beams can be of significant importance for manipulating nanoparticles and atoms especially due to local scattering force values. But the study of structured light is relevant to a detailed understanding of the dynamic properties of beams as such and to their control. 

\appendix*
\section{Estimation of the negative momenta contribution}
Below the contribution of negative momenta to the wave packet (\ref{vbet}) is evaluated. Consider the integral
\begin{equation}\label{wkcain}
{\cal I}_n=\int\limits_{\frac{1}{2\varepsilon^2}}^{\infty}dq\,e^{-q}q^{n/2}J_n\left(2\xi\sqrt{q}\right)e^{-iq\zeta},
\end{equation}
and introduce the new integration variable $u$, defined through the relation: $q=\frac{1}{2\varepsilon^2}\,u^2$. The integral (\ref{wkcain}) takes the form
\begin{equation}\label{wkkcain}
{\cal I}_n=\frac{1}{2^{n/2}\varepsilon^{n+2}}\int\limits_1^\infty du\, u^{n+1}e^{-\frac{1}{2\varepsilon^2}(1+i\zeta)u^2}J_n\Big(\frac{\sqrt{2}\xi u}{\varepsilon}\Big).
\end{equation}

For small values of $\varepsilon$ the argument of the Bessel function becomes very large (the small region close to the $\zeta$-axis will be dealt with later) and the well-known asymptotic formula \cite{gr} may be used
\begin{equation}\label{asbes}
J_n(x)\sim\sqrt{\frac{2}{\pi x}}\cos\left(x-n\,\frac{\pi}{2}-\frac{\pi}{4}\right).
\end{equation}
After substituting this expression into the integral (\ref{wkkcain}) one gets
\begin{equation}\label{poijk}
{\cal I}_n=\frac{1}{2^{n/2+3/4}\pi^{1/2}}\cdot\frac{1}{\varepsilon^{n+2}}\left[e^{-in\,\frac{\pi}{2}-i\,\frac{\pi}{4}}{\cal J}_{n+}+e^{in\,\frac{\pi}{2}+i\,\frac{\pi}{4}}{\cal J}_{n-}\right],
\end{equation}
where
\begin{eqnarray}
{\cal J}_{n\pm}&=&\int\limits_1^\infty du\, u^{n+1/2}e^{-\frac{1}{2\varepsilon^2}(1+i\zeta)u^2\pm i\frac{\sqrt{2}\xi}{\varepsilon}\,u}\nonumber\\
&=&e^{-\frac{\xi^2}{1+i\zeta}}\int\limits_1^\infty du\, u^{n+1/2}e^{-\frac{1+i\zeta}{2\varepsilon^2}\left(u\mp i\frac{\sqrt{2}\xi\varepsilon}{1+i\zeta}\right)^2}\nonumber\\
&=&e^{-\frac{\xi^2}{1+i\zeta}}{\cal K}_{n\pm}.\label{caljk}
\end{eqnarray}
Considering only leading terms in $\varepsilon$, we can write 
\begin{equation}\label{exprq}
{\cal K}_{n\pm}\sim \int\limits_1^\infty du\, u^{n+1/2}e^{-\frac{1+i\zeta}{2\varepsilon^2}u^2},
\end{equation}
and then derive a differential equation for ${\cal K}_{n\pm}$ (the symbol $\pm$ will be skipped from now on): 
\begin{eqnarray}
\frac{d {\cal K}_{n}}{d\varepsilon}&=&\frac{1+i\zeta}{\varepsilon^3}\int\limits_1^\infty du\, u^{n+5/2}e^{-\frac{1+i\zeta}{2\varepsilon^2}u^2}\nonumber\\
&=& -\frac{1}{\varepsilon}\int\limits_1^\infty du\, u^{n+3/2}\,\frac{d}{du}\,e^{-\frac{1+i\zeta}{2\varepsilon^2}u^2}\nonumber\\
&=&\frac{1}{\varepsilon}\,e^{-\frac{1+i\zeta}{2\varepsilon^2}}+\frac{n+\frac{3}{2}}{\varepsilon}\,{\cal K}_n.\label{dieqk}
\end{eqnarray}
The last expression has been obtained through integration by parts. This equation can be solved with the standard methods. The leading term turns out to be $n$-independent, and has the form
\begin{equation}\label{ledkt}
{\cal K}_n\sim \frac{1}{1+i\zeta}\,\varepsilon^2e^{-\frac{1+i\zeta}{2\varepsilon^2}},
\end{equation}
which yields the following result for ${\cal I}_n$:
\begin{equation}\label{refoi}
{\cal I}_n\sim \frac{(\pm)1}{2^{n/2+1/4}\pi^{1/2}}\cdot\frac{1}{(1+i\zeta)\varepsilon^n}\,e^{-\frac{\xi^2}{1+i\zeta}}e^{-\frac{1+i\zeta}{2\varepsilon^2}},
\end{equation}
the inessential sign being dependent on the value of $n$.

When $\xi$ is very small the estimation is straightforward, since one can use the approximation
\begin{equation}\label{asbesm}
J_n(x)\approx\frac{x^n}{2^nn!},
\end{equation}
which implies that
\begin{eqnarray}
{\cal I}_n&\approx& \frac{\xi^n}{n!}\int\limits_{\frac{1}{2\varepsilon^2}}^{\infty}dq\,e^{-q(1+i\zeta)}q^n\label{prmi}\\
&=&\frac{\xi^n}{n!(1+i\zeta)\varepsilon^{2n}}\,e^{-\frac{1+i\zeta}{2\varepsilon^2}}\big[1+O(\varepsilon^2)\big].\nonumber
\end{eqnarray}
It should be stressed that in both cases the dominating factor is $e^{-\frac{1}{2\varepsilon^2}}$.

\end{document}